\documentclass[conference]{IEEEtran}
\ifCLASSINFOpdf
\else
\fi
\usepackage{cite}

\usepackage{amsmath}

\usepackage{graphicx}

\usepackage{xspace}
\newcommand{\fig}{Fig.\xspace}
\newcommand{\tab}{Tab.\xspace}

\usepackage{multirow}

\begin{document}
%
\title{Optimal Energy Interruption Planning and Generation Re-dispatch for Improving Reliability during Contingencies}


\author{\IEEEauthorblockN{Sohail Khan, Sawsan Henein, Helfried Brunner}
	\IEEEauthorblockA{Center of Energy\\
		Austrian Institute of Technology\\
		Vienna, Austria\\
		sohail.khan@ait.ac.at, sawsan.henein@ait.ac.at, helfried.brunner@ait.ac.at}}


%


\maketitle

\begin{abstract}
Improving distribution grid reliability is a major challenge for planning and operation of distribution systems having a high share of distributed generators (DGs).
The rise of DGs share can lead to unplanned contingencies while on the other hand, they can provide flexibility in supporting grid operations after a contingency event. 
This paper presents an optimal energy interruption planning approach that dispatches the flexibility from DGs and performs a cost-optimal load shedding of flexible loads.
The proposed approach is tested on a synthetic grid, representing typical urban and low voltage feeders in EU with distribution networks modeled in radial and meshed configurations.
The study shows that the proposed optimization process can be successfully used to plan resources during contingency event and this can lead to a reduction in energy not supplied and improvement of reliability indices.
\end{abstract}


%
\IEEEpeerreviewmaketitle

\section{Introduction}
The power systems globally are experiencing a rise in power consumption. 
Along with it, the share of environment-friendly generation in the form of solar and wind energy is rising.
Such generation sources are generally connected to the distribution networks and are termed as distributed generators (DGs)~\cite{gao2014reliability}.
As the share of the DGs increase, it is anticipated that local generation can reduce the energy losses and may lead to reliability improvements due to the distributed nature and close proximity to the loads. 
However, they can also increase the likelihood of failures, for instance, line over-loading due to reverse power, voltage constraint violations, and transformer overloading.
Therefore, it is imperative to assess the impacts of DGs on the reliability of the network during contingency studies~\cite{xu2016reliability}. 

After contingency occurrence, the power system may experience no problem if it is able to re-balance or a severe problem due to t e outage of one element or a very critical situation due to cascade outages of several grid elements
Remedial actions are planned by the system operator against contingency events, that may involve disconnection of part of loads in the network generally termed as load shedding, operate switches to enable alternate supply points for de-energized loads or re-dispatch DGs in the network \cite{mishra2012contingency}.
As the share of DGs in the network and system load increase, there is more likelihood of the contingency events.
Furthermore, the load shedding and re-dispatch flexibility of supporting DGs need to be optimally planned during contingency events in order to reduce the interruption costs and decrease the overall energy not supplied\cite{al2018contingency}.
This demonstrates the importance of finding the optimal generation and load scheduling which allows the network operators to minimize the energy interrupted per customers and reduce power curtailments and meanwhile reduces the costs of load shedding and increase customers satisfaction~\cite{chandramohan2010operating}.

In recent years, several studies have considered techniques for locating, sizing and operation of DG units installed in the distribution systems to improve system performance in contingency situations.
Several studies related to DGs have introduced algorithms for loss minimization, improve voltage profile~\cite{leghap1ploss,prakash2016optimal,buaklee2013optimal}, improving the reliability~\cite{ghadimi2013method, borges2006optimal,tolba2018impact,banerjee2011reliability,conti2014probability,trebolle2010reliability} and placement of fault indicators or switch devices~\cite{billinton1996optimal,wang2008reliability,ho2010optimal,etemadi2008distribution}.
In addition, the potential of the energy storage system in improving the reliability and reduction of energy interruption costs has been discussed in~\cite{saboori2015reliability}.
However, few studies have taken an integrated approach to the optimization of DG re-dispatch and load shedding flexibility during contingency scenarios. 
Authors in~\cite{yassami2011optimal} have used a heuristic optimization approach in which optimal energy interruption planning is performed while the DG and load flexibility are optimized. 
In comparison, in this paper, a deterministic approach is taken to solve a similar optimization problem. 
The presented work is part of the H2020 project called INTERPLAN.
The goal of the project is to provide an INTEgrated opeRation PLANning tool towards the pan-European network, to support the European Union in reaching the expected low-carbon targets while maintaining the network security\cite{graditi2019critical}. 
The proposed method can be used by system operators as an offline planning tool to analyse the expected contingencies to help them react to future outages by using pre-planned optimized loads and generation schedules
It is also anticipated that this process will need to be repeated by the system operator as the critical nodes in the system change, thus a deterministic formulation optimization approach is required and is presented in the paper. 

The paper is organized as, Sec.\ref{sec:modeling} discusses the contingency list formulation, flexibility modeling, the optimization problem formulation, and reliability parameters calculation.
The second part of the paper discusses the use-case in Sec.\ref{sec:use_case} where the proposed optimization process is applied to the system and the results are compared on the basis of reliability indices. Finally, the conclusion is discussed in Sec.\ref{sec:conclusion}.

\section{Modeling}\label{sec:modeling}
These critical nodes and lines in a distribution network can be determined by the proximity of their operational state to the network constraint limits, such as bus voltage thresholds or line loading limits.
The risk associated with critical nodes is not only dependent on the loading but also on the condition that may deteriorate along with time and can lead to the disconnection. 
This risk is captured by the voltage margin between the peak voltage and the voltage limits and similarly for line loading and their operational limits.
During the course of the study, a set of critical nodes is determined by either running a load flow at one point in time or for a duration spanning a day or year depending on the availability of data.
Afterwards, the line loading and voltage limits of all nodes in the system are monitored.
The criteria in terms of voltage and line loading is set and the subsequent results analysis based on it defines the set of the critical contingencies. 

\begin{figure} [htb]
	\centering
	\includegraphics[trim=0cm 0cm 0cm 0cm, width=0.48\textwidth]{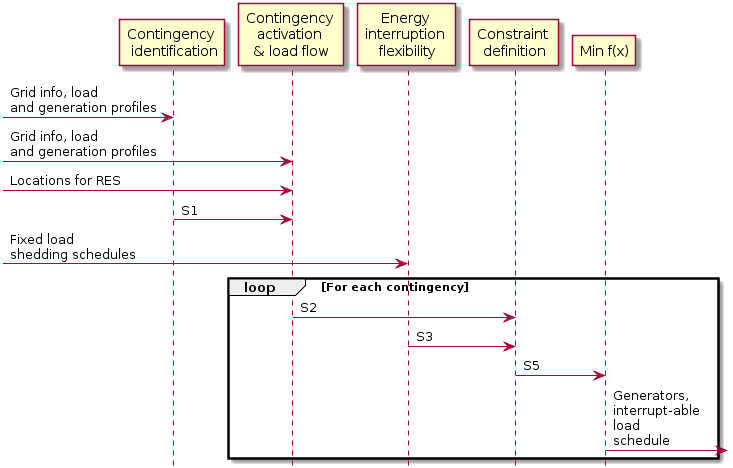}
	\caption{Sequence diagram for the optimal energy interruption planning.}
	\label{fig:sequence_diagram}
\end{figure}

The optimal energy interruption planning as proposed in this paper is performed for a contingency event.
Thus in the planning phase, an iterative process is performed that activates a contingency in the grid and perform the optimization.
The post-contingency state of the system can lead to loads being not served or change in the operational state of the network that can lead to network constraint violation.
During which, some of the loads may loose their primary energy supply in case if there is no alternate source of power.
Considering the availability of an alternate source of supply, either the loads shedding can be performed or generation resources can be re-dispatched to ensure that power is supplied to critical loads.
The resources in the post-contingency state of the system need to be optimally dispatched in order to reduce the energy interruption costs.
The state flow diagram of proposed energy interruption planning is shown in \fig~\ref{fig:sequence_diagram}.
The first stage is the identification of critical contingencies. During which, load flow studies are performed based on the grid information and load \& generation profiles.
Afterward, a contingency from the set is activated in the network and load flow is performed.
The load shedding flexibility and associated costs are defined by the operator.
It along with the generation re-dispatch flexibility are formulated as a constraint matrix for the optimization problem. 

\subsection{Modeling of flexibility}\label{sec:modeling_flexibility}
The active power of each load $P_L$  has a part that can be shed $P_{LS}$ after the contingency event.
While the cost associated with load shedding $c_{LS}$ acts as a multiplier to $P_{LS}$ in the objective function. 
The load shedding can be a continuous value or can be specified as steps. 
In this case, they are modeled as ramp rate constraints. 
The generator set-points $P_G$ and their re-dispatch capacity $P_{GD}$ provides an additional dimension in flexibility with associated cost given as, $c_{GS}$.
It is assumed that if a contingency creates an island in the system, then the generator posses sufficient grid forming capabilities.

\subsection{Optimization problem formulation}\label{sec:optimization_problem}
The optimal energy interruption problem is formulated using Newton-Lagrange method. 
This corresponds to the minimization of the Lagrange function~\cite{salehi2012implementation, digsilentpf2019manual} given as:
\begin{equation}
	L\left(\vec{x},\vec{s},\vec{\lambda}
	\right)
	=
	f(\vec{x}) - 
	\mu \sum_{i} \log (s_i) + \\
	\vec{\lambda^T}
	\left[\;
	g(\vec{x}) + h(\vec{x}) + \vec{s}\;
	\right]
\end{equation}
where, $\vec{x}$ is the state vector at $i^{th}$ bus, $\left(P^i, Q^i, v^i, \theta^i\right)$.
\begin{flalign}
g\left(\vec{x}\right) & = 0\; , \quad \textrm{load flow equations}\\
h\left(\vec{x}\right) & \leq 0\; . \quad \textrm{inequality constraints, e.g., $v^i \leq 1.1$ p.u.}
\enspace . \nonumber
\end{flalign}

Rest of the variables includes: $\vec{s}$ is the slack variable for each inequality constraint, where, $\vec{s}\ge 0$ such that $h(\vec{x})+\vec{s}=0$,
$\vec{\lambda}$ as Lagrangian multiplier and
$\mu$ is the multiplier for logarithmic function of $\vec{s}$ as a penalty factor.
The objective function, $f(\vec{x})$ is given as:
\begin{equation}
	f\left(\vec{x}\right) = 
	w_L\sum_{j=1}^{N_{L}}
	c_{LS}^{j}
	\left|
	P_{LS}^{j}
	\right| + 
	w_G\sum_{k=1}^{N_{G}}
	c_{GD}^{k}
	\left|
	P_{GD}^{k}
	\right|
\end{equation}
here, $c_{LS}^{j}$ and $c_{GD}^{k}$ are the cost factor of load curtailed and generator re-dispatch, $w_L$ and $w_G$ are the weighing factors of loads and generators and $N_L$ \& $N_G$ are number of loads that can shed and number of generators that can be re-dispatched. 

The outcomes of the energy interruption optimization process are set-points for load shedding and re-scheduled operating points of active and reactive power of dispatch-able generators. 
Reliability of the system is assessed by the system indices such as system average interruption frequency index (SAIFI) with units, [1/C/a], system average interruption duration index (SAIDI) with units, [h/a], energy not supplied (ENS) with units, [MW] and others. 
Reliability analysis is performed after activation of each contingency and the parameters are calculated. 
The process is repeated after the operating points from the optimization process are activated in the system.
The comparison provides and insight in to the potential reliability improvement due to the proposed approach.

\subsection{Reliability calculation}\label{sec:reliability_calculation}
The reliability indices are calculated as in~\cite{schiesser2009reliability}.
The SAIFI reliability index indicates how often the average customer experiences a sustained interruption per year. It is calculated as~\cite{6209381},
\begin{equation}
\textrm{SAIFI} = \frac{\sum_{i}^{N_L}\textrm{ACIF}_i\cdot C_i}{\sum_{i}^{N_L}C_i}
\;\; \textrm{in} \;\;
[1/C/a] \; ,
\end{equation}
where average customer interruption frequency (ACIF) is given as,
\begin{equation}
\textrm{ACIF}_i = \sum_{k}^{N_C}\textrm{Fr}_k\cdot \frac{P_{LS}^{i,k}}{P_{L}^{i}}
\;\; \textrm{in} \;\;
[1/a] \; .
\end{equation}
Here, $C_i$ is the number of customers supplied by the load point $i$, $\textrm{Fr}_k$ is the frequency of occurrence of contingency $k$ and $P_{LS}^{i,k}$ is the amount of the load lost at $i^{\textrm{th}}$ load point and for $k^{\textrm{th}}$ contingency.
Similarly, SAIDI is calculated as,
\begin{equation}
\textrm{SAIDI} = \frac{\sum_{i}^{N_L}\textrm{ACIT}_i\cdot C_i}{\sum_{i}^{N_L}C_i}
\;\; \textrm{in} \;\;
[h/C/a] \; ,
\end{equation}
where average customer interruption time (ACIT) is given as,
\begin{equation}
\textrm{ACIT} = \sum_{k}^{N_C} 8760 \cdot \textrm{Pr}_k\cdot 
\frac{P_{LS}^{i,k}}{P_{L}^{i}}
\;\; \textrm{in} \;\;
[h/a] \; .
\end{equation}
Here, $\textrm{Pr}_k$ is the probability of occurrence of contingency $k$.
In addition to the frequency/expectancy indices, the energy index considered is ENS, and is given as,
\begin{equation}
\textrm{ENS} = \sum_{i}\textrm{LPENS}_i 
\;\; \textrm{in} \;\; 
[\textrm{MWh}/a] \; ,
\end{equation}
\begin{equation}
\textrm{LPENS}_i  = \textrm{ACIT}_i \cdot \left(\widehat{P_{LS}^{i}} + \widehat{P_{LNS}^{i}}\right)
\;\; \textrm{in} \;\;
[\textrm{MWh}/a] \; .
\end{equation}
where, $\widehat{P_{LS}^{i}}$ and $\widehat{P_{LNS}^{i}}$  are the normalized average amount of power shed and disconnected at load point $i$ for all the contingencies.

\section{Use-case}\label{sec:use_case}
\begin{figure} [htb]
	\centering
	\includegraphics[trim=0cm 0cm 0cm 0cm, width=0.5\textwidth]{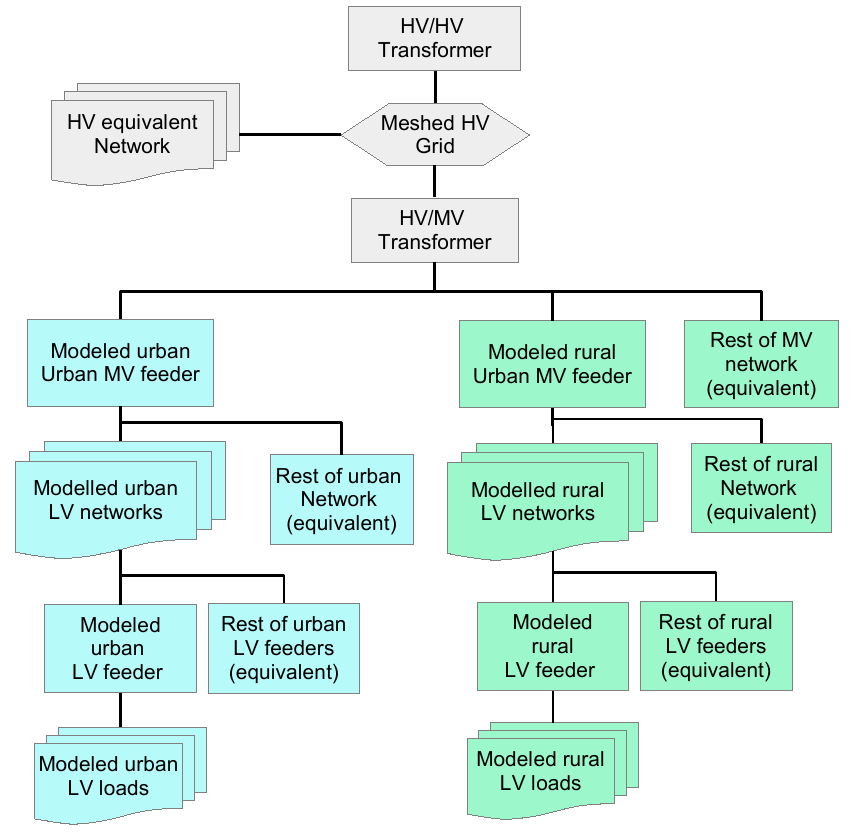}
	\caption{Sythetic grid overview diagram.}
	\label{fig:test_case}
\end{figure}

The test case considered is a synthetic grid formulated based on network equivalence and represents the HV, MV and LV sections, the overview diagram is shown in \fig~\ref{fig:test_case}.
The network has two formulations. In the first formulation, the modeled urban and rural LV feeders have radial structure and in the second formulation, the end points of the feeders are connected together separately for urban and rural networks forming a meshed topology. The load flow data of both formulations are provided in~\cite{khan_sohail_2019_3379046} in \textit{pypower} format with visualizations obtained using Pandapower~\cite{thurner2018pandapower}. 
The optimization is performed in~DIgSILENT PowerFactory\footnote{\texttt{https://www.digsilent.de/en/powerfactory.html}}.

\begin{figure} [htb]
	\centering
	\includegraphics[trim=0cm 0.1cm 0cm 0.2cm, clip, width=0.49\textwidth]{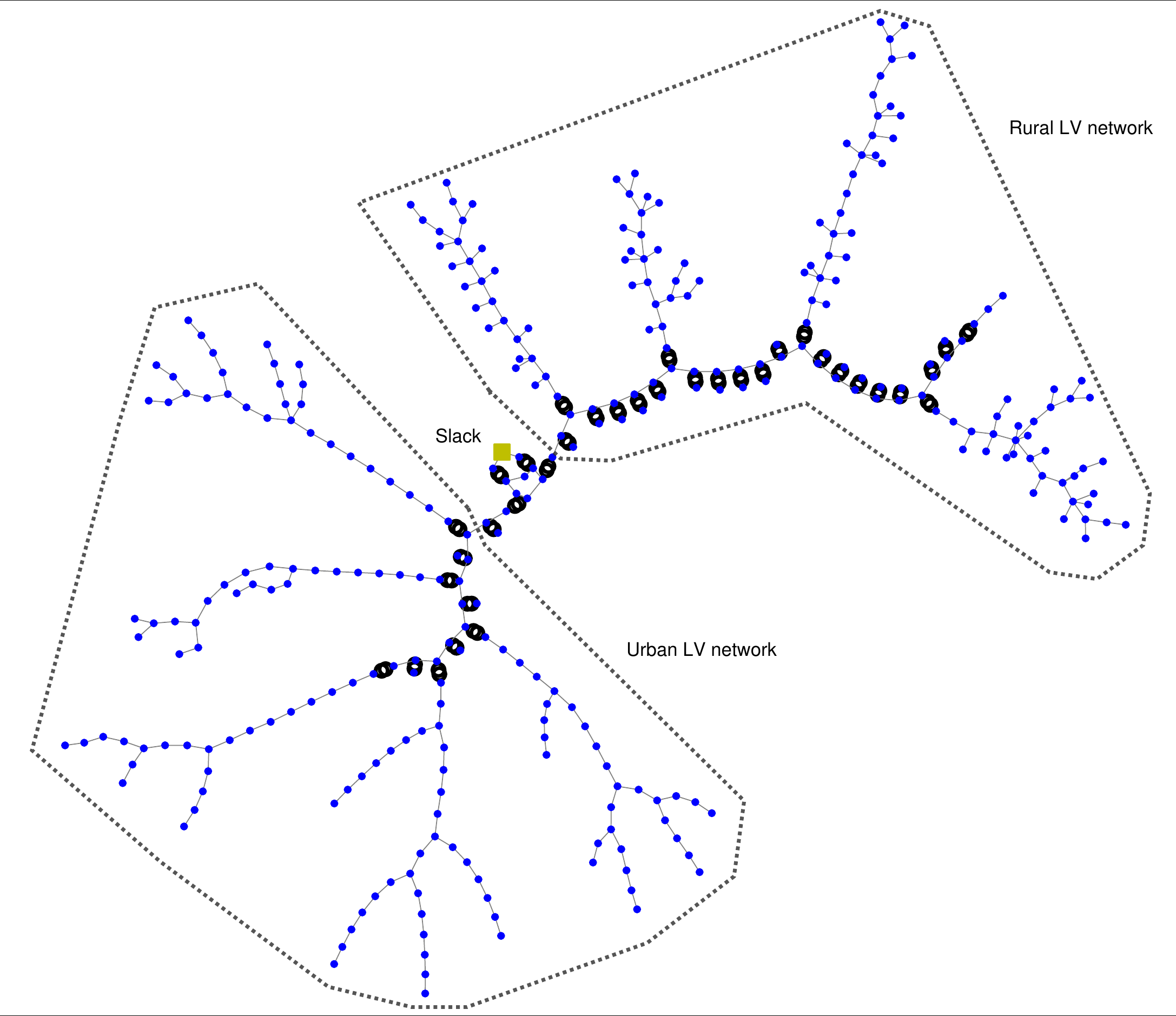}
	\caption{Test network with radial LV networks.}
	\label{fig:test_case_radial_pandapower}
\end{figure}

\begin{figure} [htb]
	\centering
	\includegraphics[trim=0.1cm 0cm 0cm 0cm, clip, width=0.49\textwidth]{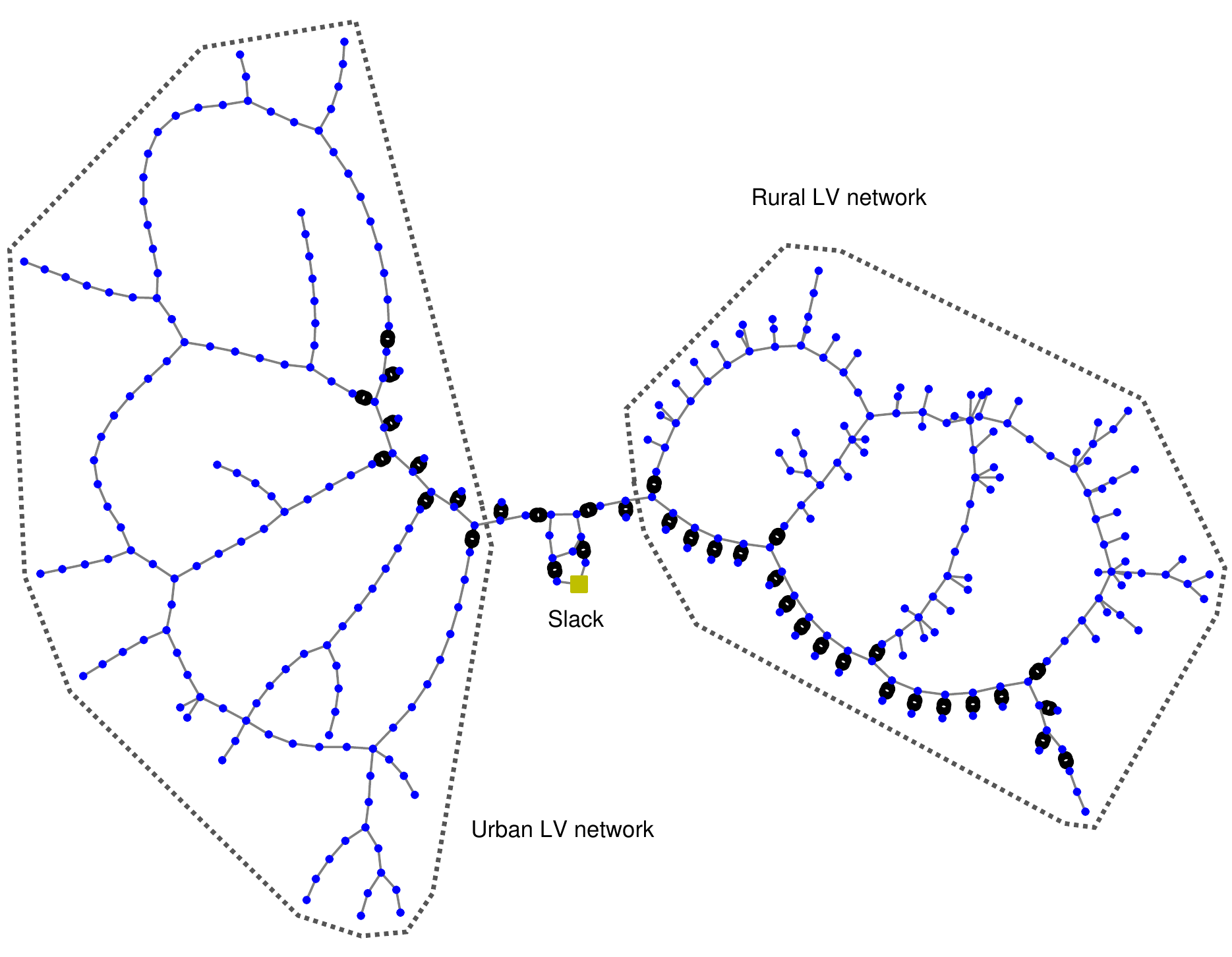}
	\caption{Test network with meshed LV networks.}
	\label{fig:test_case_meshed_pandapower}
\end{figure}

The synthetic grid represent a network covering HV, MV and LV voltage levels. The MV networks are modeled in two categories, rural and urban feeders. Five LV feeders in each category are modeled in detail, while rest of the LV feeders are represented by equivalent loads. The network is designed from the perspective of power networks commonly found in EU.
The network has two topological scenarios, the radial and meshed formulation as shown in \fig~\ref{fig:test_case_radial_pandapower} and ~\ref{fig:test_case_meshed_pandapower} , respectively.

\subsection{Reliability data of the test system}\label{sec:reliability_data_test_system}
The reliability data for the network is shown in \tab~\ref{table:reliability_parameters}. It is obtained from~\cite{transmission2003ieee,billington1984reliability}.
The outage cost is taken as a linear graph between energy and costs $(\textrm{MWh},  \; \$/\textrm{kWh})$. It spans between $(0,9.2)$ and  $(3,20)$ and is linearly extrapolated.
The stochastic model of generation is based on three derated states level and is shown in \tab~\ref{table:stochastic_generation_model}.
\begin{table}[htb]
	\def\arraystretch{1.3}\tabcolsep=4pt
	\centering
	\caption{Reliability parameters.}
	\label{table:reliability_parameters}
\begin{tabular}{l|l|c|c|c|}
	\hline
	\multicolumn{2}{|c|}{\multirow{2}{*}{\textbf{\begin{tabular}[c]{@{}c@{}}Reliability\\ model\\ parameters\end{tabular}}}} & \textbf{\begin{tabular}[c]{@{}c@{}}Failure\\ rate\end{tabular}} & \textbf{\begin{tabular}[c]{@{}c@{}}Additional\\ failure \\ rate per\\ connection\end{tabular}} & \textbf{\begin{tabular}[c]{@{}c@{}}Repair\\ time\end{tabular}} \\ \cline{3-5} 
	\multicolumn{2}{|c|}{}                                                                                                   & $\lambda \; [1/a]$                                              & $\lambda\; [1/a]$                                                                              & $d \; [h]$                                                     \\ \hline
	\multicolumn{1}{|l|}{\multirow{3}{*}{\begin{tabular}[c]{@{}l@{}}Busbar\\ failure\end{tabular}}}         & 11 kV          & 0.001                                                           & 0.001                                                                                          & 2                                                              \\ \cline{2-5} 
	\multicolumn{1}{|l|}{}                                                                                  & 33 kV          & 0.001                                                           & 0.001                                                                                          & 2                                                              \\ \cline{2-5} 
	\multicolumn{1}{|l|}{}                                                                                  & 230 kV         & 0.22                                                            & 0.22                                                                                           & 10                                                             \\ \hline
	\multicolumn{2}{l|}{}                                                                                                    & $\lambda \; [1/(a*km)]$                                         & -                                                                                              & $d \; [h]$                                                     \\ \hline
	\multicolumn{1}{|l|}{\multirow{3}{*}{\begin{tabular}[c]{@{}l@{}}Line\\ failure\end{tabular}}}           & 11 kV          & 0.065                                                           & -                                                                                              & 5                                                              \\ \cline{2-5} 
	\multicolumn{1}{|l|}{}                                                                                  & 33 kV          & 0.046                                                           & -                                                                                              & 8                                                              \\ \cline{2-5} 
	\multicolumn{1}{|l|}{}                                                                                  & 230 kV         & 0.02                                                            & -                                                                                              & 10                                                             \\ \hline
	\multicolumn{2}{l|}{}                                                                                                    & $\lambda \; [1/a]$                                              & $\lambda \; [1/a]$                                                                             & $d \; [h]$                                                     \\ \hline
	\multicolumn{1}{|l|}{\multirow{3}{*}{\begin{tabular}[c]{@{}l@{}}Transformer\\ failure\end{tabular}}}    & 11kV/0.4kV     & 0.015                                                           & -                                                                                              & 200                                                            \\ \cline{2-5} 
	\multicolumn{1}{|l|}{}                                                                                  & 138kV/33kV     & 0.01                                                            & -                                                                                              & 15                                                             \\ \cline{2-5} 
	\multicolumn{1}{|l|}{}                                                                                  & 230kV/138kV    & 0.02                                                            & -                                                                                              & 768.                                                           \\ \hline
\end{tabular}
\end{table}
\begin{table}[htb]
	\def\arraystretch{1.3}\tabcolsep=3.5pt
	\centering
	\caption{Stochastic generation model.}
	\label{table:stochastic_generation_model}
	\begin{tabular}{|l|l|l|l|l|l|}
		\hline
		\textbf{State} & \multicolumn{1}{c|}{\textbf{\begin{tabular}[c]{@{}c@{}}Availability \\ \%\end{tabular}}} & \multicolumn{1}{c|}{\textbf{\begin{tabular}[c]{@{}c@{}}Probability \\ \%\end{tabular}}} & \multicolumn{1}{c|}{\textbf{\begin{tabular}[c]{@{}c@{}}Duration \\ h\end{tabular}}} & \multicolumn{1}{c|}{\textbf{\begin{tabular}[c]{@{}c@{}}Frequency \\ 1/a\end{tabular}}} & \multicolumn{1}{c|}{\textbf{\begin{tabular}[c]{@{}c@{}}Duration \\ h/a\end{tabular}}} \\ \hline
		State 1        & 100                                                                                      & 85.54                                                                                   & 10                                                                                  & 749.33                                                                                 & 7493                                                                                  \\ \hline
		State 2        & 80                                                                                       & 13.45                                                                                   & 44.72                                                                               & 26.35                                                                                  & 1178                                                                                  \\ \hline
		State 3        & 0                                                                                        & 1.01                                                                                    & 74                                                                                  & 1.2                                                                                    & 88.5                                                                                  \\ \hline
	\end{tabular}
\end{table}

\subsection{Application of the method to the use-case}\label{sec:application_on_use_case}
The test system is simulated with a number of contingencies defined on the critical nodes and lines in the network.
For each contingency, the energy interruption optimization as discussed in Sec.\ref{sec:optimization_problem} is performed that leads to optimal load shedding and generator re-dispatch to satisfy the network constraints.
System reliability indices are calculated before and after the optimization process. 
The comparison of cases with and with-out optimizations for the radial topology of the test network is shown in \tab~\ref{table:radial_topology_results}.
It can be observed that optimal interruption planning reduced the SAIFI and SAIDI indices, the major improvement has been on the ENS, where the index is improved by 32\%.
The rows 5 to 8 represents the average values of load shedding, losses and active power dispatched over all the contingencies.
It can be observed that about 17\% of the active power capacity is re-dispatched.
\begin{table}[htb]
	\def\arraystretch{1.2}\tabcolsep=6pt
	\centering
	\caption{Simulation results of optimal energy interruption planning for test network with radial topology}
	\label{table:radial_topology_results}
	\begin{tabular}{lllll}
		& Unit  & Base case & \begin{tabular}[c]{@{}l@{}}Energy \\ interruption \\ optimization\end{tabular} & \begin{tabular}[c]{@{}l@{}}Percentage \\ difference\end{tabular} \\ \hline
		SAIFI                                                                 & 1/C/a & 0.0116    & 0.0115                                                                               & -1.4609                                                             \\ \hline
		SAIDI                                                                 & h/C/a & 0.0276    & 0.0272                                                                               & -1.3117                                                             \\ \hline
		ENS                                                                   & MW    & 0.0183    & 0.0123                                                                               & -32.8311                                                            \\ \hline
		EIC                                                                   &  M\$/a     & 0.0003    & 0.0002                                                                               & -33.1577                                                            \\ \hline
		Load shedding                                                         & MW    & 0         & -0.0049                                                                              & 0                                                                   \\ \hline
		Active power loss                                                     & MW    & 0.1049    & 0.1036                                                                               & -1.1865                                                             \\ \hline
		Reactive power loss                                                   & Mvar  & -7.0531   & -7.0865                                                                              & 0.4735                                                              \\ \hline
		\begin{tabular}[c]{@{}l@{}}Active power \\ dispatched\end{tabular} & MW    & 3.8612    & 4.5315                                                                               & 17.3599 \\ \hline                                                           
	\end{tabular}
\end{table}

The results on the meshed topology of the network are shown in \tab~\ref{table:meshed_topology_results}. Comparing the results of radial and meshed topology, it can be observed that the proposed approach leads to more improvement for the meshed network due to more degrees of freedom attributed to an alternate source of energy supply points.
The load shedding requirements are also decreased due to meshed topology.
Furthermore, the average value of the active power dispatched is also reduced.
\begin{table}[htb]
	\def\arraystretch{1.2}\tabcolsep=6pt
	\centering
	\caption{Simulation results of optimal energy interruption planning for test network with meshed topology}
	\label{table:meshed_topology_results}
	\begin{tabular}{lllll}
		& Unit  & Base case & \begin{tabular}[c]{@{}l@{}}Energy\\ interruption\\ optimization\end{tabular} & \begin{tabular}[c]{@{}l@{}}Percentage\\ difference\end{tabular} \\ \hline
		SAIFI                                                             & 1/C/a & 0.0004    & 0.0004                                                                       & -2.7473                                                         \\ \hline
		SAIDI                                                             & h/C/a & 0.0008    & 0.0008                                                                       & -2.7473                                                         \\ \hline
		ENS                                                               & MW    & 0.0017    & 0.0011                                                                       & -34.7291                                                        \\ \hline
		EIC                                                               & M\$/a     & 0         & 0                                                                            & -34.7552                                                        \\ \hline
		Load shedding                                                     & MW    & 0         & -0.0042                                                                      & 0                                                               \\ \hline
		Active power loss                                                 & MW    & 0.1048    & 0.1043                                                                       & -0.5005                                                         \\ \hline
		Reactive power loss                                               & Mvar  & -7.0994   & -7.122                                                                       & 0.3194                                                          \\ \hline
		\begin{tabular}[c]{@{}l@{}}Active power\\ dispatched\end{tabular} & MW    & 4.6042    & 5.1867                                                                       & 12.6509 \\ \hline                                                       
	\end{tabular}
\end{table}

The comparison shows marked improvement in reliability indices due to energy interruption planning.
The important challenge while implementing this approach is to ensure the network observability of the system after the contingency event. In addition, to effectively perform the remedial operation, the controls signal communication with the loads and/or generators that are re-dispatched needs to be ensured. 

\section{Conclusion}\label{sec:conclusion}
This paper presents an optimal energy interruption planning process that utilizes the load shedding and generation re-dispatch flexibility to satisfy the network constraints after a contingency event.
The optimal results represent the minimum cost of flexibility activation. 
The proposed approach can be used by the network planners to evaluate technological options in the contingency planning process. However, further studies are required to study the impact of remedial actions on the congestion in the network and to implement congestion avoidance as a criterion in the planning process.

\section*{Acknowledgment}
The research work  leading to  the developed  results was fully supported by  the European Union’s Horizon 2020 Research and Innovation Programme under Grant Agreement No. 773708.

\bibliographystyle{IEEEtran}

\end{document}